\documentclass[11pt]{article}
\usepackage[utf8]{inputenc}
\usepackage{amssymb,amsmath}
\usepackage[hidelinks]{hyperref}
\usepackage[a4paper,margin=2.5cm]{geometry}

\DeclareMathOperator{\FST}{FST}

\def\R{\mathbb{R}}
\def\diffd{\mathrm{d}}

\title{Exact solution and precise asymptotics of a Fisher-KPP type front}

\author{Julien
Berestycki\footnote{\href{mailto:Julien.Berestycki@stats.ox.ac.uk}{\ttfamily Julien.Berestycki@stats.ox.ac.uk}, Department
of Statistics, University of Oxford, UK}, \'Eric
Brunet\footnote{\href{mailto:Eric.Brunet@lps.ens.fr}{\ttfamily Eric.Brunet@lps.ens.fr}, Laboratoire de
Physique
Statistique, \'Ecole Normale
Sup\'erieure, PSL Research University; Universit\'e Paris Diderot Sorbonne
Paris-Cit\'e; Sorbonne Universit\'es UPMC Univ Paris 06; CNRS.},
Bernard
Derrida\footnote{\href{mailto:Bernard.Derrida@lps.ens.fr}{\ttfamily Bernard.Derrida@lps.ens.fr}, Laboratoire de
Physique
Statistique, \'Ecole Normale
Sup\'erieure, Coll\`ege de France.}}

\begin{document}

\maketitle

\begin{abstract}

The present work concerns a version of the Fisher-KPP equation where the
nonlinear term is replaced by a saturation mechanism, yielding a free
boundary problem with  mixed conditions. Following an idea proposed in
\cite{BrunetDerrida.2015}, we show that the Laplace transform of the
initial condition is directly related to some functional of the front
position $\mu_t$. We then obtain precise asymptotics of the front position
by means of singularity analysis. In particular, we recover the so-called
Ebert and van Saarloos correction \cite{EbertvanSaarloos.2000}, we obtain
an additional term of order $\log t /t$ in this expansion, and we give
precise conditions on the initial condition for those terms to be present. 

\end{abstract}
\section{Introduction}
Since the seminal works of Fisher \cite{Fisher.1937} and Kolmogorov,
Petrovsky, Piscounov \cite{KPP.1937}, travelling
wave equations such as the Fisher-KPP equation
\begin{equation}
\partial_t H = \partial_x^2 H + H - H^2\qquad\text{(Fisher-KPP)},
\label{FKPP}
\end{equation}
have attracted uninterrupted attention
\cite{Bramson.1978, Bramson.1983, McKean.1975, EbertvanSaarloos.2000}.
Although a lot is understood on the long time behaviour of
these equations, explicit calculations are usually difficult
because of the non-linearities.
Recently \cite{Henderson.2016,BBHR.2016,BrunetDerrida.2015}, it has been
shown that some front equations with the non-linearities replaced by
a saturation at some moving boundary exhibit the same long time asymptotic
as the Fisher-KPP equation \eqref{FKPP}.

In this paper, we  study such a linearised Fisher-KPP equation 
which generalises \cite{Henderson.2016,BBHR.2016}, and we analyse it by
extending to the continuum the approach developed in
\cite{BrunetDerrida.2015} for a lattice version of the problem.
We consider the joint evolution
\begin{equation}
\left\{\begin{array}{l}
\partial_t h = \partial_x^2h+h\quad\text{if $x>\mu_t$},\\
h(\mu_t,t)=\alpha,\\
\partial_x h(\mu_t,t)=\beta,
\end{array}\right.\label{ours}\end{equation}
of a boundary $\mu_t$  and of a function $h(x,t)$ defined for $x>\mu_t$
with a given initial condition $h_0(x)$ defined for $x\ge0$.
The parameters $\alpha$ and $\beta$ are fixed.
This is a free boundary problem
\cite{HilhorstHulshof.1991,BraunerHulshofLunardi.2000, Freidlin.1985,
Friedman.2010} where
both $h(x,t)$ and $\mu_t$ are unknown quantities to be
determined. We limit our discussion to
\begin{equation}
\begin{cases}
\alpha>0,\\\beta\in\R,\end{cases}\qquad\text{or}\qquad\begin{cases}\alpha=0,\\\beta>0,\end{cases}
\label{ab}
\end{equation}
because the other cases can be obtained by changing the sign of $h_0$.
By analogy with the Fisher-KPP equation where most studies focus on
positive fronts, we also assume that
\begin{equation}
h_0(x)\ge0.
\end{equation}
In this case, the solution $h(x,t)$, whenever it exists, remains positive
for all $t>0$ and $x>\mu_t$.

One way to think about \eqref{ours} is to first choose
\textit{a priori} a smooth boundary $t\mapsto \mu_t$ with $\mu_0=0$, and
then to solve the system $\big\{\partial_t h = \partial_x^2h+h$ if
$x>\mu_t$ and $h(\mu_t,t)=\alpha\big\}$ with initial condition $h_0$, as in
\cite{Henderson.2016,BBHR.2016}.
Then comes the main difference from \cite{Henderson.2016,BBHR.2016}: out of
all the possible choices for the trajectory $t\mapsto\mu_t$ of the boundary, we
select (whenever it exists and is unique)
the one for which the solution $h$ satisfies
$\partial_xh(\mu_t,t)=\beta$ at all times $t>0$.

Whether there exists a solution $(\mu_t, h)$ to \eqref{ours} for an
arbitrary initial condition $h_0$  and whether such a solution is unique
are not  easy questions (see discussion in Section~\ref{sec:rem}). However,
it is easy to show (see Section~\ref{sec:TW})  that \eqref{ours} admits,
for $v$ large enough, positive travelling wave solutions, \textit{i.e.}
solutions of the form $\mu_t=vt$ and $h(x,t)=\omega_v(x-vt)$ for $x>vt$,
where $\omega_v(z)\ge0$ and $\omega_v(\infty)=0$. For suitable choices of
$h_0$, one could expect the solution to \eqref{ours} to converge to one of
these positive travelling waves, in a  way similar
to the Fisher-KPP equation, in the sense that
\begin{equation}
h(\mu_t+z,t)\xrightarrow[t\to\infty]{}\omega_v(z)\quad\text{for $z>0$},
\qquad \frac{\mu_t}t\to v.
\label{ifcvg}
\end{equation}
This, however, is not always the case. For example, we show in
Section~\ref{sec:rem} that \eqref{ifcvg} does not hold in the ($\alpha=0$,
$\beta=1$) case when $\int\diffd z\,h_0(z)\ne1$. See also \cite{HamelNadin.2012}
for an example in the Fisher-KPP case.

The key of  our approach in the present paper is  the following
exact relation between $\mu_t$ and the initial condition $h_0$: 
when \eqref{ifcvg} holds, 
for any $r$ 
such that both sides converge,
\begin{equation}
\int_0^\infty\diffd z\, h_0(z)e^{r z} 
=-\frac\alpha r +\Big(\beta+\frac\alpha r\Big)
\int_0^\infty \diffd t\, e^{r\mu_t-(1+r^2)t}.
\label{main}
\end{equation}
This equation is a continuous version of a result obtained for a 
system defined on the lattice \cite{BrunetDerrida.2015}.

As in \cite{BrunetDerrida.2015}, one can 
use~\eqref{main} to analyse how the large time asymptotics of $\mu_t$
depends on
the initial condition $h_0$.
For $\alpha+\beta>0$ and when \eqref{ifcvg} holds we obtain that
\begin{subequations}
\label{mainresults}
\begin{align}
\mu_t&= 2t - \frac32\log t + \text{Cst} +  o(1)
&&\text{iff }
\int_0^\infty\diffd x\, h_0(x) x e^x <\infty,
\label{Bram}\\
\mu_t&=
2t - \frac32\log t + \text{Cst} 
- 
3\dfrac{\sqrt\pi}{\sqrt t} + o(t^{-1/2}) &&\text{if }
\displaystyle
\int_0^\infty\diffd x\, h_0(x) x^2 e^x <\infty,\label{rvS}
\\
\mu_t&=2t-\frac32\log t+\text{Cst}-3\frac{\sqrt\pi}{\sqrt t}+
		\frac98(5-6\log2)\frac{\log t}t +\mathcal
O\Big(\frac1t\Big)
	&&\text{if $\int_0^\infty\diffd x\, h_0(x)x^3 e^x<\infty$}. \label{Eric}
\end{align}
\end{subequations}
(Here, and everywhere in this paper, ``Cst'' stands for some constant term.)
Observe that 
\eqref{Bram} coincides with  Bramson's result for the position of a Fisher-KPP front 
\cite{Bramson.1978,Bramson.1983}, and that  \eqref{rvS} reproduces the
prediction of Ebert and van Saarloos (recently proved  in \cite{NRR.2016} for
compactly supported initial conditions). 
This raises the question of whether the $\frac{\log t}t$ term in \eqref{Eric}
should be present for other Fisher-KPP like equations.
The sufficient conditions on $h_0$ in \eqref{rvS} and \eqref{Eric} are
close to be necessary; the precise necessary conditions are given
in Section~\ref{sec:asres}.

More detailed asymptotics of $\mu_t$ for other initial conditions
are given in Section~\ref{sec:asres}. In particular, we argue that when
$\alpha+\beta<0$, the solution to~\eqref{ours} behaves as a ``pushed
front'' \cite{vanSaarloos.2003} rather than a Fisher-KPP or ``pulled'' front when
$\alpha+\beta>0$. (The case $\alpha+\beta=0$ would correspond to some
critical situation between the ``pulled'' and ``pushed'' cases.)

The long time asymptotics \eqref{mainresults} are the
same  as in the lattice version considered in \cite{BrunetDerrida.2015}:
\begin{equation}
\partial_t h(n,t) = \begin{cases}h(n,t)+h(n-1,t)&\text{if $h(n,t)<1$},
\\ 
0 &\text{otherwise}.
\end{cases}
\label{eqBD15}
\end{equation}
The approach
used in \cite{BrunetDerrida.2015} relied on a relation similar to 
\eqref{main} which was, however,  more complicated and limited to 
the ``pulled''  case.

The particular ($\alpha=0$, $\beta=1$) case is related to a problem
discussed in the mathematical literature
\cite{Henderson.2016,BBHR.2016}, where the question was
how to choose
$\mu_t$ for a given $h_0$ in such a way that the solution to
\begin{equation}
\left\{\begin{array}{l}
\partial_t h = \partial_x^2h+h\quad\text{if $x>\mu_t$},\\
h(\mu_t,t)=0,\\
\end{array}\right.\label{oldBC}\end{equation}
converges to a travelling wave: $h(\mu_t+z,t)\to\omega_v(z)>0$.
It was shown that \eqref{Bram} must hold for a fast decaying initial
condition $h_0$ (see also \cite{HamelNolenRoquejoffreRyzhik.2013}).
By requiring furthermore that the convergence of
$h(\mu_t+z,t)$ to $\omega_v(z)$ is fast,
\cite{Henderson.2016,BBHR.2016} obtain results compatible with \eqref{rvS}.

A linear equation with a moving Dirichlet condition similar to \eqref{ours}
is expected to appear as
the hydrodynamic limit of the $N$-BBM: consider a Branching
Brownian Motion where $N$ particles diffuse and branch independently. At
each branching event, the leftmost particle is removed so that the
population size~$N$ is kept constant. Call $\mu_t$ the position of the leftmost
particle and $h(x,t)$ the empirical density of particles divided by $N$.
Then, as $N\to\infty$, the evolution of $h$ and $\mu_t$ becomes
deterministic and satisfy
\cite{CarinciDeMasiGiardinaPresutti.2016, 
DeMasiFerrariPresuttiSporano-Loto.2017}
the following system of equations:
\begin{equation}
\left\{\begin{array}{l}
\partial_t h = \partial_x^2h+h\quad\text{if $x>\mu_t$},\\
h(\mu_t,t)=0,
\\[1ex]\displaystyle
\int_{\mu_t}^\infty \diffd x\, h(x,t)=1,
\end{array}\right.\label{NBBM}\end{equation}
where both $h$ and $\mu_t$ are unknown quantities to be solved for.
(See also \cite{DurrettRemenik.2011} who obtain the analogue of
\eqref{NBBM} for a specific $N$-Branching Random Walk.)
By differentiating the last line of  \eqref{NBBM} with respect to
$t$, one gets $0=-h(\mu_t,t)+\int_{\mu_t}^\infty\diffd
x\,(\partial_x^2h+h)=-h(\mu_t,t)-\partial_xh(\mu_t,t)+\int_{\mu_t}^\infty\diffd
x\,h$. With the conditions in \eqref{NBBM}, one obtains that $\partial_xh(\mu_t,t)=1$ for all $t$.
Therefore, \eqref{NBBM} reduces to our problem~\eqref{ours}  with
($\alpha=0$, $\beta=1$).

\smallskip

The structure of this paper is the following:
\begin{itemize}
\item In Section~\ref{sec:TW}, we write the travelling wave solutions to
\eqref{ours} and discuss their similarity with the travelling waves of
the Fisher-KPP equation when
$\alpha+\beta>0$.

\item In Section~\ref{sec:magic}, we establish the main
relation~\eqref{main}, on which the analysis of the long time asymptotics
of $\mu_t$ are based.

\item In Section~\ref{sec:rem},
we discuss for which initial conditions one can expect a travelling wave
solution \eqref{ifcvg}.

\item In Section~\ref{sec:asres} we obtain, from a singularity
analysis of \eqref{main}, 
precise asymptotics for the position $\mu_t$ of the front
in the ``pulled'' case ($\alpha+\beta>0$), in the ``pushed'' case
($\alpha+\beta<0$) and in the ``critical'' case ($\alpha+\beta=0$).

\item In Section~\ref{sec:Munier}, we briefly describe the long time
behaviour of $h(x,t)$ and of $\mu_t$ when the front does not converge to
a travelling wave. This allows to recover a recent result
on a self-consistent method to find the typical position of the rightmost
particle in a BBM \cite{MuellerMunier.2014}.

\end{itemize}

\section{Travelling waves}\label{sec:TW}

It is easy to determine all travelling wave solutions of \eqref{ours}.
Writing
\begin{equation}
\mu_t = v t,\qquad h(x,t)=\omega_v(x-vt)\quad \text{if $x>v t$,}
\label{formTW}
\end{equation}
one finds that $\omega_v$ satisfies
\begin{equation}
\omega_v''+v\omega_v'+\omega_v = 0\quad\text{for $z>0$},\qquad
\omega_v(0)=\alpha,\qquad\omega_v'(0)=\beta.
\end{equation}
With the extra condition that $\omega_v(z)$ goes to zero as $z\to+\infty$, a solution
only exists for $v>0$.
Writing 
\begin{equation}
v=\gamma+\frac1\gamma,
\end{equation}
this gives
\begin{equation}
\omega_v(z)=\frac{(\alpha+\beta\gamma)e^{-\gamma z}-\gamma(\alpha\gamma+\beta)e^{-\frac1\gamma
z}}{1-\gamma^2}\text{ for $v\ne2$},\qquad
\omega_2(z)=[\alpha+(\alpha+\beta)z]e^{-z}.
\label{TW}
\end{equation}

For $v<2$, the exponential rates $\gamma$ and $\gamma^{-1}$ are complex
conjugates, and the travelling wave decays to zero with oscillations
around zero. Such a travelling wave cannot be reached by a non-negative initial condition. 

For $v>2$, the exponential rates $\gamma$ and $\gamma^{-1}$ are real, one
smaller than 1 and the other larger than 1. We always choose $\gamma$ such
that $0<\gamma\le 1\le \gamma^{-1}$.

For $v=2$, the exponential decay rate is $\gamma=1$, with a $z$ in the
prefactor.

\smallskip 

Recalling \eqref{ab}, there are three subcases to be considered:
\begin{itemize}
\item If $\alpha+\beta>0$.
All the travelling waves \eqref{TW} for $v\ge2$ remain positive for all $z>0$. They
decay like $e^{-\gamma z}$ as $z\to\infty$ 
for $v>2$ and like $z e^{-z}$ for $v=2$. This is
very similar to what is known for the Fisher-KPP case, which is often
called the ``pulled'' case \cite{vanSaarloos.2003}.
\item If $\alpha>0$ and $\alpha+\beta<0$.
The travelling waves remain positive if and only if $v\ge v_*$ where
\begin{equation}
v_*=\gamma_*+\frac1{\gamma_*}\quad\text{with
}\gamma_*=\frac{\alpha}{-\beta}\in(0,1).
\end{equation}
For $v>v_*$,  the travelling waves decay like $Ae^{-\gamma z}$ with
$A>0$. For
$v\in(2,v_*)$ they also decay like $Ae^{-\gamma z}$, but with $A<0$.
For $v=v_*$, 
the travelling wave is simply equal to $\alpha \exp(-\gamma_*^{-1}z)$;
the decay is much faster than for any other velocity. This situation is
sometimes called the ``pushed case''. \cite{vanSaarloos.2003}
\item If $\alpha>0$ and $\alpha+\beta=0$. This case is critical
between the pushed and the pulled case. All the travelling waves for
$v\ge2$ are positive, but the travelling wave for $v=2$ decays as $e^{-z}$
instead of $ze^{-z}$.
\end{itemize}

\section{Derivation of the main relation (\ref{main})} \label{sec:magic}

In this section, we establish the relation \eqref{main} between the initial
condition $h_0$ and the position $\mu_t$ of the solution to \eqref{ours}.
With $h(x,t)$ the solution to \eqref{ours}, let us introduce

\begin{equation}
g(r,t)=\int_0^\infty\diffd z\, h(\mu_t+z,t)e^{rz},
\label{defg}
\end{equation}
where $r$ is real small enough for the integral to converge. ($r$ can be
negative if needed. We do not discuss here initial conditions $h_0$ that 
increase so fast that no such $r$ exists.)
Differentiating with respect to $t$ and replacing $\partial_t h$ by its
expression $\partial_t h = \partial_x^2h+h$, one gets
\begin{equation}
\partial_t g(r,t)=\int_0^\infty\diffd z\, \Big[ \dot\mu_t\partial_x
h(\mu_t+z,t)+\partial_x^2h(\mu_t+z,t)+h(\mu_t+z,t)\Big]e^{rz}.
\label{15}
\end{equation}
Integration by parts with $h(\mu_t,t)=\alpha$ and $\partial_x
h(\mu_t,t)=\beta$ yields
\begin{equation}
\int_0^\infty\diffd z\,\partial_x
h(\mu_t+z,t) e^{rz}=-\alpha -r g(r,t),\qquad
\int_0^\infty\diffd z\,\partial_x^2
h(\mu_t+z,t) e^{rz}=\alpha r +r^2 g(r,t)-\beta.
\end{equation}
Then
\begin{equation}
\partial_t g(r,t)=\big[r^2-r\dot\mu_t  +1\big] g(r,t) +\alpha
r -\beta-\alpha\dot\mu_t.
\end{equation}
This can of course be integrated:
\begin{equation}
g(r,t)= -\frac\alpha r +\bigg[C_r-\bigg(\beta+\frac\alpha r\bigg)\int_0^t
\diffd s \,e^{-(r^2+1)s+r\mu_s}\bigg] e^{(r^2+1)t - r \mu_t},
\label{25b}
\end{equation}
with $C_r$ a constant of integration. By taking the limit $t\to0^+$,
\begin{equation}
C_r=\frac\alpha r + \int_0^\infty\diffd z\,h_0(z)e^{rz}.
\label{Cr0}
\end{equation}

Up to now, we made no assumption on the long time behaviour of the solution $h$ of \eqref{ours},
and \eqref{25b} with \eqref{Cr0} is always valid as long as the solution
exists. From now on, we  restrict ourselves to the case where the solution
of \eqref{ours}
converges to a travelling wave with some velocity
$v=\lim_{t\to\infty}\frac{\mu_t}t$, as in \eqref{ifcvg}.
As in Section~\ref{sec:TW}, using that $v\ge2$, we write 
\begin{equation}
v=\gamma+\frac1\gamma\quad\text{with $\gamma\in(0,1]$},
\end{equation}
It is then clear that $\exp\big[(r^2+1)t-r\mu_t\big]$ in \eqref{25b}
diverges as $t\to\infty$ for
$r<\gamma$ or $r>\gamma^{-1}$, and goes to zero for
$r\in(\gamma,\gamma^{-1})$. Furthermore, the left hand side $g(r,t)$ of
\eqref{25b} converges when $t\to\infty$ to $\int\diffd
z\,\omega_v(z)e^{rz}$ which is finite
when $r<\gamma$, as can be checked with \eqref{TW}
from the large $z$ behaviour of the travelling waves $\omega_v$. 

Then, for  $r<\gamma$, the left hand side of \eqref{25b} converges
and the outer exponential in the right hand side diverges, so
the expression in square brackets must vanish in the $t\to\infty$ limit:
\begin{equation}
C_r=\bigg(\beta+\frac\alpha r\bigg)\int_0^\infty
\diffd s \,e^{-(r^2+1)s+r\mu_s} \qquad\text{for $r<\gamma$}.
\label{Crinfty}
\end{equation}
Combining \eqref{Cr0} and \eqref{Crinfty} leads to
our main relation~\eqref{main}.

A similar equation was obtained in \cite{DewynneHowisonOckendonXie.1989}
for the Stefan problem, which is another free boundary problem.

\section{Some remarks on the solutions to (\ref{ours})} \label{sec:rem}

In this section, we give some conditions on $h_0$ for the solution of
\eqref{ours} to converge to a travelling wave. 

By an explicit construction of the solution, we first argue that in the
($\alpha=1$, $\beta=0$) case, the solution to \eqref{ours} converges to
a travelling wave for a large class of initial conditions $h_0$; this class
includes all the decreasing functions smaller than~1  which decay
exponentially fast at infinity. In fact, as explained in
Section~\ref{const}, the construction works as long as $h(x,t)$ remains
below $1$ for all $t>0$ and $x>\mu_t$.

Then, we show in Section~\ref{equiv} that one can relate the solution of
\eqref{ours} for arbitrary ($\alpha$, $\beta$) and
initial condition $h_0$  to a solution of the same problem \eqref{ours} but
with
parameters $(\alpha=1,\beta=0)$ and an initial condition $\eta_0$ computed
from $h_0$. This mapping leads to a necessary condition for the front to converge to a travelling wave:
\begin{equation}
\beta\le0\qquad\text{or\qquad}\int_0^\infty\diffd z\, h_0(z)e^{-\frac\alpha\beta z}=\beta.
\label{condM0}
\end{equation}
It turns out, in the ($\alpha=0$, $\beta>0$) case, that this is
also sufficient as $\eta_0$ is then in fact decreasing, as shown below
in \eqref{eta0h0}. On the other hand,
when \eqref{condM0} fails, the solution cannot converge to
a travelling wave;  consider $g(-\alpha/\beta,t)$ from \eqref{25b}, with
the value of $C_{r=-\alpha/\beta}$ given by \eqref{Cr0}:
\begin{equation}
g(-\alpha/\beta,t)=\beta+ \left(\int_0^\infty\diffd
z\,h_0(z)e^{-\frac\alpha\beta z}-\beta\right)
e^{\big[\big(\frac\alpha\beta\big)^2+1\big]t+\frac\alpha\beta
\mu_t}.
\label{gsp}
\end{equation}
When \eqref{condM0} is not satisfied $\beta>0$ and the expression in
parenthesis is non zero. If the front were to reach a travelling wave, the
exponential factor in the right-hand-side of \eqref{gsp} would diverge, in
contradiction with the fact that the left-hand-side would converge to
$\int\diffd
z\,\omega_v(z)e^{-(\alpha/\beta)z}$, as can be seen from the definition
\eqref{defg} of $g$. 

Finally in Section \ref{4.3}, we analyse how the asymptotic decay of
$h_0$ determines that of $\eta_0$. In certain regimes, $\eta_0$ decays more
slowly than $h_0$ and this shows that \eqref{ours} does no longer behave as the
usual KPP front.  

The mapping of Section~\ref{equiv} gives some insight on the unicity of the
solutions to \eqref{ours}. Consider two given  boundaries $t\mapsto
\mu_t^{+}$
and $t\mapsto \mu_t^{-}$ and let $h^{+}$ and $h^{-}$ be the solutions
to $\partial_t h^{\pm}=\partial_x^2h^{\pm}+h^{\pm}$ for $x>\mu_t^{\pm}$
with
$h^{\pm}(\mu_t^{\pm},t)=0$ and $h^{\pm}(x,0)=h_0(x)$ given. Then, if
$\mu_s^+\ge\mu_s^-$ for all $s\le t$ with a strict equality on some
interval, it is clear that
$\int_{\mu_t^+}^\infty \diffd x\,h^+(x,t) <\int_{\mu_t^-}^\infty \diffd
x\,h^-(x,t)$. This suggests that the solution to \eqref{NBBM} is unique.
(There exists a rigourous proof in some cases, see
\cite{DeMasiFerrariPresuttiSporano-Loto.2017}.) But \eqref{NBBM} is the
same as \eqref{ours} with ($\alpha=0$, $\beta=1$), and through the  mapping
of Section~\ref{equiv}, the solution to the general $(\alpha$, $\beta$)
case must be unique.

\subsection{The case \texorpdfstring{($\alpha=1$, $\beta=0$)}{(alpha=1,
beta=0)}}
\label{const}

Let us show how to construct the solution to \eqref{ours} with
parameters ($\alpha=1$, $\beta=0$) and an initial condition $h_0(x)$
defined for $x>0$. As usual, we assume that $h_0 \in[0,1]$
decays exponentially fast.

For $n>1$, we introduce $H_n(x,t)$ as the solution to
\begin{equation}
\partial_t H_n
=\partial_x^2H_n+H_n-H_n^n,\qquad
H_n(x,0)= \begin{cases}h_0(x)&\text{if $x\ge0$},\\1&\text{if
$x<0$}.\end{cases}
\label{Hn}
\end{equation}
By standard comparison principle, one has for any $x$ and $t$,
\begin{equation}
0\le H_n(x,t) \le H_{m}(x,t)\le1\quad\text{if $n\le m$}.
\end{equation} 
One concludes that, for fixed $x$ and $t$, the large $n$ limit of $H_n$
exists, and we define
\begin{equation}
H(x,t):=\lim_{n\to\infty} H_n(x,t).
\label{HHn}
\end{equation}
Clearly, $0\le H(x,t)\le1$.

Assume for now that $h_0$ is a decreasing function of $x$. 
By standard results on Fisher-KPP equations \cite{Bramson.1983}, all of the 
$H_n(x,t)$ are non-increasing in $x$, and so is $H$ after taking the limit.
Therefore there exists a  $\mu_t$ such that
\begin{equation}
\begin{cases}	H(x,t) =1 & \text{if $x\le\mu_t$}\\
		H(x,t) <1 & \text{if $x>\mu_t$}.	
\end{cases}
\label{mutexists}
\end{equation}
The position $\mu_t$ above must be finite for all $t>0$. Indeed,  the
function $H$ cannot be uniformly equal to 1 since it must be smaller than
the solution $L$ to the linearised equation $\partial_t L =\partial^2_x
L+L$ with the same initial condition, and $L$ is smaller than 1 for $x$
large enough. Similarly, $H$ cannot be everywhere smaller than 1:
if it were, $H$ would be equal to $L$, and this is impossible because
$L>1$ for $x$ negative enough.

The couple $(\mu_t,H)$ is thus the solution to the system \eqref{ours} with
parameters $(\alpha=1,\beta=0)$ and initial condition $h_0$.
The condition that $h_0$ is decreasing is convenient, but by no mean
necessary; what really matters to identify $H$ with the solution $h$ to
\eqref{ours} is that \eqref{mutexists} holds.

By standard Fisher-KPP theory \cite{Bramson.1983}, each of the $H_n$ in
\eqref{Hn} converges as $t\to\infty$ to some travelling wave with
a velocity~$v$ which depends on $h_0$ but not on $n$. This $v$ is also the
velocity of the front described by the linearised equation $\partial_t
L =\partial_x^2L+L$. The bounds $H_n(x,t)\le H(x,t)\le L(x,t) $ thus yield
that asymptotically the front $H$ must also have the velocity $v$.
Although this does not directly yields the convergence towards a
travelling wave, it is nevertheless a very strong indication that such a
convergence holds. Indeed, establishing the asymptotic velocity is usually
the first step of proving the convergence to a travelling wave in many
reaction diffusion equations, see for instance the celebrated result of
Aronson and Weinberger \cite{AronsonWeinberger.1975} for the Fisher-KPP
equation.

\subsection{Mapping the general \texorpdfstring{($\alpha$,
$\beta$)}{(alpha, beta)} case into the
\texorpdfstring{($\alpha=1$, $\beta=0$)}{(alpha=1, beta=0)} case} \label{equiv}

We present a procedure to transform the general ($\alpha$, $\beta$) case
of  \eqref{ours}
with initial condition $h_0$
into the $(\alpha=1,\beta=0)$ one. We start by defining 
\begin{equation}
\eta_0(x)=\begin{cases} \displaystyle {e^{\frac\alpha\beta x}}\left[1-\frac1\beta\int_0^x \diffd z\,
e^{-\frac\alpha\beta z} h_0(z)\right] \qquad &\text{ if } \beta\neq
0,\\[2ex] h_0(x) /\alpha  \qquad &\text{ if } \beta= 0.\end{cases}
\label{eta0h0}
\end{equation}
Let $\eta(x,t)$ be the solution to \eqref{ours} with
($\alpha=1$, $\beta=0$)
and initial condition $\eta_0(x)$:
\begin{equation}
\partial_t\eta = \partial_x^2\eta+\eta\text{ if $x>\mu_t$},\qquad
\eta(\mu_t,t)=1,\qquad \partial_x\eta(\mu_t,t)=0,\qquad
\eta(x,0)=\eta_0(x).
\label{ours10}
\end{equation}
Then $h(x,t)$ defined as
\begin{equation}
h(x,t)= \alpha\eta(x,t)-\beta\partial_x\eta(x,t) 
\label{etatoh}
\end{equation}
is solution of \eqref{ours} with parameters $(\alpha,\beta)$ and initial condition $h_0$. 
Both the $h$ front and the $\eta$ front have
the same boundary $\mu_t$.

When $\beta=0$, the result is trivial by linearity.
When $\beta\neq 0$ we need to check that:
\begin{itemize}
\item $h$ is solution to $\partial_t h = \partial_x^2h+h$ for $x>\mu_t$.\\
This is obvious from \eqref{etatoh} and \eqref{ours10} by linearity.
\item $h(\mu_t,t)=\alpha$.\\
This is also obvious from \eqref{etatoh} and \eqref{ours10}.
\item $\partial_x h(\mu_t,t)=\beta$.\\
This one is less obvious. It works from \eqref{etatoh} because one has
\begin{equation}
\partial_x^2\eta(\mu_t,t)=-1\qquad\text{for all $t>0$},
\label{eta''}
\end{equation}
as can be seen by taking the time derivative of
$1=\eta(\mu_t,t)$; one gets 
$0=\dot\mu_t\partial_x\eta(\mu_t,t) + \partial_x^2\eta(\mu_t,t)
+ \eta(\mu_t,t) = \partial_x^2\eta(\mu_t,t)+1$.
\item $h(x,0)= h_0(x)$.\\
Indeed, it follows
from \eqref{eta0h0} that $\alpha\eta_0-\beta\eta_0'=h_0$.
\item If $\beta \neq 0$, one needs the condition that $\eta_0(x)$ is continuous and such that $\eta_0(0)=1$.\\
The fact that the condition holds is obvious from \eqref{eta0h0}. Here is
why it is needed: if the condition did not hold, then
around any discontinuity point in $\eta_0$ (or around
$\mu_t$ if $\eta_0(0)\ne1$), the solution $\eta(x,t)$ would have
 an arbitrarily
large slope at short times even though it would be continuous.
This would mean in \eqref{etatoh} that $h(x,t)$
would be unbounded around the problematic points in $\eta_0$ and,
therefore, would not be an acceptable solution to \eqref{ours}. 
\end{itemize}

With the procedure  \eqref{eta0h0}, \eqref{ours10}, \eqref{etatoh}, it is
clear that $h(x,t)$ converges to a travelling wave if and only if
$\eta(x,t)$ does.  The first thing to check is whether the $\eta_0(x)$ in
\eqref{eta0h0} goes to zero exponentially fast as $x\to\infty$. When
$\beta\le0$, this is always the case. When $\beta>0$, one checks easily  that this is the case if and
only if the term in square brackets in \eqref{eta0h0} goes to zero as
$x\to\infty$. (In either case, one needs to use the hypothesis that
$h_0$ decays exponentially fast.) We have therefore justified the criterion \eqref{condM0}.

In particular, in the case $\alpha=0$ and $\beta>0$ and for a $h_0$ such that \eqref{condM0}
holds, the initial condition $\eta_0$ \eqref{eta0h0} 
is a decreasing function, which is sufficient to ensure that the front reaches a travelling wave.

\subsection{Asymptotics of \texorpdfstring{$\eta_0$}{eta0}} \label{4.3}

It is interesting to compare the large $x$ behaviours of
$\eta_0$ (the initial condition of the ($\alpha=1$,
$\beta=0$) problem) and of $h_0$
(the initial condition of the original ($\alpha$,
$\beta$) problem).
For simplicity, we limit the discussion to initial conditions
of the form
\begin{equation}
h_0(x)\sim A x^\nu e^{-\gamma x}\quad\text{as $x\to\infty$},
\label{h0typical}
\end{equation}
for some values of $A>0$, $\gamma>0$ and $\nu$. With this form,
one finds in
\eqref{eta0h0} 
\begin{equation}
\eta_0(x)\sim  \begin{cases}
\dfrac A{\alpha+\gamma\beta}x^\nu e^{-\gamma x}&\text{if $\beta>0$ and
assuming \eqref{condM0}, or if $\beta<0$ and
$\gamma<-\dfrac\alpha\beta$}
,\\[2.5ex]
\text{[some constant]} e^{\frac\alpha\beta x} &\text{if $\beta<0$ and $\int\diffd
z \, e^{-\frac\alpha\beta z}h_0(z)<\infty$}
,\\[2.5ex]
\dfrac A {-\beta}(\log x) e^{\frac\alpha\beta x}&\text{if 
$\beta<0$ and $\gamma=-\dfrac\alpha\beta$ and $\nu=-1$}
,\\[2.5ex]
\dfrac A{-\beta(\nu+1)}
x^{\nu+1} e^{\frac\alpha\beta x} &\text{if 
$\beta<0$ and $\gamma=-\dfrac\alpha\beta$ and $\nu>-1$}
.
\end{cases}
\label{pulledtopushed}
\end{equation}

From \eqref{pulledtopushed}, we will obtain in the next section the
asymptotic behaviour of $\mu_t$ for front in the
so-called ``pushed'' regime (if $\alpha+\beta<0$) by translating it into
the ``pulled'' problem ($\alpha=1$, $\beta=0$) and the initial condition
(with different asymptotic) given by \eqref{pulledtopushed}.

\section{The position of the front} \label{sec:asres}
\def\cst{a}
\def\cstknown{a}

In this section, we use the main relation~\eqref{main} to relate the
long time asymptotics of the 
position $\mu_t$ of the front to the initial condition $h_0$, assuming that
the solution converges to a travelling wave.

When $\beta>-\alpha$, 
we find for initial
conditions that decay fast enough that, as $t\to\infty$,
\begin{equation}
\begin{aligned}
\mu_t&=2t-\frac32\log t+\cst+o(1) &&
	\text{iff $\int\diffd x\, h_0(x)x e^x<\infty$},\\
\mu_t&=2t-\frac32\log t+\cst-3\frac{\sqrt\pi}{\sqrt t}+
		o\Big(\frac1{\sqrt t}\Big) &&
	\text{if $\int\diffd x\, h_0(x)x^2 e^x<\infty$},\\
\mu_t&=2t-\frac32\log t+\cst-3\frac{\sqrt\pi}{\sqrt t}+
		\frac98(5-6\log2)\frac{\log t}t +
o\Big(\frac{\log t}t\Big) &&
	\text{if $\int\diffd x\, h_0(x)x^3 e^x<\infty$},
\end{aligned}
\label{pull1}
\end{equation}
where we have no expression of $\cst$ as a function of $h_0$, $\alpha$ and
$\beta$.
The asymptotics are the same as for the Fisher-KPP equation or any ``pulled
front''. We recover in particular the Bramson term $-\frac32\log t$,
the Ebert and van Saarloos correction $-3\sqrt\pi /\sqrt t$ and a new
universal term in $(\log t)/t$. Precise necessary conditions for the two
last lines of \eqref{pull1} are given in \eqref{truecond1} and
\eqref{truecond2}.

If the initial condition decays as
\begin{equation}
h_0(x)\sim  A x^\nu e^{-\gamma x}\quad\text{as $x\to\infty$},
\label{againh0}
\end{equation}
we also find that
\begin{equation}
\begin{aligned}
\mu_t &= \Big(\gamma+\frac1\gamma\Big)t +\frac\nu\gamma\log
t+\cstknown+o(1) && \text{if $\gamma<1$}
,\\
\mu_t&=2t-\frac{1-\nu}2\log t +\cstknown+o(1)
  && \text{if $\gamma=1$ and $\nu\in(-2,\infty)$}
,\\
\mu_t&=2t-\frac32\log t +\log\log t +\cstknown+o(1)
  && \text{if $\gamma=1$ and $\nu=-2$}
,\\
\mu_t&=2t-\frac32\log t +\cst-b t^{1+\frac\nu2}
   +o\big(t^{1+\frac\nu2}) && \text{if $\gamma=1$ and $\nu\in[-3,-2)$}
,\\
\mu_t&=2t-\frac32\log t +\cst-3\frac{\sqrt\pi}{\sqrt t}
  - b t^{1+\frac\nu2} +o\big(t^{1+\frac\nu2}) && \text{if $\gamma=1$ and $\nu\in(-4,-3)$}
,\\
\mu_t&=2t-\frac32\log t +\cst-3\frac{\sqrt\pi}{\sqrt t}
  + b\frac{\log t}t +o\Big(\frac{\log t}t\Big)
 && \text{if $\gamma=1$ and $\nu=-4$}
.
\end{aligned}
\label{pull2}
\end{equation}
For the first three lines, there is a relatively
simple expression of $a$ as a function of  $\alpha$, $\beta$, $A$,
$\gamma$, $\nu$.
The constant $b$ can
be expressed as a function of $a$, $\nu$, $\alpha$, $\beta$ and $A$.
All these expressions are compatible with what is already known for the Fisher-KPP
equation \eqref{FKPP}.

The results \eqref{pull1} and \eqref{pull2} concern
pulled front equation ($\beta>-\alpha$). For pushed and critical
fronts, we could also use the main relation \eqref{main} to derive the
asymptotic position of the front. It is however simpler to translate the
front into a pulled front, as explained in Section~\ref{4.3}, and use
\eqref{pulledtopushed} to obtain the results. We find that
\begin{itemize}
\item When $\beta<-\alpha$ (``pushed'' front equation), setting
$\gamma_*=\alpha/(-\beta)$ and $v_*=\gamma_*+\gamma_*^{-1}$,
\begin{equation}
\begin{aligned}
\mu_t &= \Big(\gamma+\frac1\gamma\Big)t +\frac\nu\gamma\log
t+a+o(1) && \text{for \eqref{againh0} if $\gamma<\gamma_*$}
,\\
\mu_t&=v_*t+\frac{\nu+1}{\gamma_*}\log t +a+o(1)
  && \text{for \eqref{againh0} if $\gamma=\gamma_*$ and $\nu\in(-1,\infty)$}
,\\
\mu_t&=v_*t+\frac1{\gamma_*}\log\log t +a+o(1)
&& \text{for \eqref{againh0} if $\gamma=\gamma_*$ and $\nu=-1$}
,\\
\mu_t&=v_* t +a+o(1)&&\text{if $\int\diffd x\,h_0(x)e^{\gamma_*
x}<\infty$}
.
\end{aligned}
\label{pushedresults}
\end{equation}
The constant $a$ depends on $\alpha$, $\beta$ and the whole function $h_0$
in the last line. For the other three lines, $a$ can be expressed as
a function of $\alpha$, $\beta$, $A$, $\gamma$, $\nu$.
{To illustrate how \eqref{pushedresults} is obtained, consider the second line
(\textit{i.e.}
$h_0\sim A x^\nu e^{-\gamma_* x}$ with $\nu>-1$): the last line of
\eqref{pulledtopushed} tells us that this corresponds to a pulled front
with initial condition $\eta_0\sim A' x^{\nu+1} e^{-\gamma_* x}$, and the
first line of \eqref{pull2} with $\nu$ replaced to $\nu+1$ gives the
answer. For the third line of \eqref{pushedresults}, the initial
condition would be $\eta_0\sim A'
(\log x)e^{-\gamma_*x}$, which is not in \eqref{pull2}, but which could
be computed easily using the methods explained in this section.}
\item When $\beta=-\alpha$ (critical front equation), we get
\begin{equation}
\begin{aligned}
\mu_t &= \Big(\gamma+\frac1\gamma\Big)t +\frac\nu\gamma\log
t+\cst+o(1) && \text{for \eqref{againh0} if $\gamma<1$}
,\\
\mu_t&=2t+\frac{\nu}{2}\log t +\cst+o(1)
  && \text{for \eqref{againh0} if $\gamma=1$ and $\nu\in(-1,\infty)$}
,\\
\mu_t&=2t-\frac12\log t +\log\log t +\cst+o(1)
&& \text{for \eqref{againh0} if $\gamma=1$ and $\nu=-1$}
,\\
\mu_t&=2 t -\frac12\log t +\cst+o(1)&&\text{if $\int\diffd x\,h_0(x)e^{
x}<\infty$}
.
\end{aligned}
\label{criticalresults}\end{equation}
As in the pushed case,
the constant $a$ depends on $\alpha$, $\beta$ and the whole function $h_0$
in the last line. For the other three lines, $a$ can be expressed as
a function of $\alpha$, $\beta$, $A$, $\gamma$, $\nu$.
\end{itemize}

We now turn to the derivation of \eqref{pull1} and \eqref{pull2}. From
now on in this section, we
assume that $\beta>-\alpha$ (``pulled'' case) and that the front
converges to a travelling wave with velocity $v$:
\begin{equation}
h(\mu_t+z,t)\to\omega_v(z),\qquad\frac{\mu_t}t\to v,\qquad
v=\lambda+\frac1\lambda\ \text{with $\lambda\le1$.}
\label{deflambda}
\end{equation}
We introduce
\begin{equation}
\Psi_1(r)=\int_0^\infty\diffd z\, h_0(z)e^{r z}  
,\qquad
\Psi_2(r)=\int_0^\infty \diffd t\, e^{r\mu_t-(1+r^2)t}
,\qquad
\gamma=\sup\big\{r; \Psi_1(r)<\infty\big\}
.
\label{defPsi}
\end{equation}
($\gamma$ is the exponential decay rate of $h_0$.
By hypothesis on $h_0$, one has $\gamma>0$.)

$\Psi_1(r)$ is finite for $r<\gamma$.
$\Psi_2(r)$ is finite for $r<\lambda$ or $r>\lambda^{-1}$, 
where $\lambda$ is defined in \eqref{deflambda}.
Our main relation~\eqref{main} states that
\begin{equation}
\Psi_1(r) =-\frac\alpha r +\Big(\beta+\frac\alpha r\Big) \Psi_2(r)
\qquad
\text{for $r<\min(\lambda,\gamma)$.}
\label{magic2}
\end{equation}

The main idea to obtain the asymptotics of $\mu_t$ is to express that both
sides of \eqref{magic2} have the same first singularity in $r$. Matching 
the
position of this singularity determines the velocity of the front, see
Section~\ref{velsec}, while matching the nature of the singularity
determines the sublinear term, as explained in the subsequent sections.

\subsection{Velocity selection}
\label{velsec}

The basic idea to find the final velocity of the front is that the two
functions $\Psi_1(r)$ and $\Psi_2(r)$ become singular at the same
value of $r$. 
Given $\gamma$ (a property of initial condition $h_0$, see \eqref{defPsi}),
we want to compute the velocity $v$ or, equivalently, $\lambda$, 
see \eqref{deflambda}.

It can be checked that
\begin{itemize}
\item $\Psi_1(r)$ is analytic for any $r<\gamma$ but becomes singular
at $r=\gamma$ when $\gamma$ is finite,

\item $\Psi_2(r)$ is analytic for any $r<\lambda$. When $\lambda<1$, it
becomes singular
at $r=\lambda$.
(For $\lambda=1$, we will see that, depending on $\mu_t$,
$\Psi_2(r)$ can be either singular or analytic at $r=\lambda=1$.)
\end{itemize}

Then, \eqref{magic2} implies that
\begin{equation}
\lambda=\begin{cases} \gamma & \text{if $\gamma\le1$\quad(both $\Psi_1$ and
$\Psi_2$ are singular at $r=\lambda=\gamma$)},\\
1&\text{if $\gamma>1$\quad(both $\Psi_1$ and
$\Psi_2$ are analytic at $r=1$)}.
\end{cases}
\end{equation}
The velocity is then $v=2$ if $\gamma\ge1$ and
$v=\gamma+\gamma^{-1}$ if $\gamma\le1$.

Remark: in the ``pushed case'', the prefactor $(\beta+\frac\alpha r)$  of $\Psi_2$ in
\eqref{magic2} vanishes at $r=\gamma_*=\alpha/(-\beta)$ and, when $\gamma>\gamma_*$, 
$\Psi_1$ is analytic at $r=\gamma_*$ while $\Psi_2$ has a single pole.

\subsection{The singularities in \texorpdfstring{$\Psi_1$ and
$\Psi_2$}{Psi1 and Psi2}} 
\label{sing1}

\subsubsection*{The singularity in $\Psi_1$}

When the initial condition $h_0$ is of the form~\eqref{againh0}, one has
\begin{equation}
\Psi_1(\gamma-\epsilon) = \int_0^\infty \diffd z\,\big(h_0(z)e^{\gamma z}\big)\,e^{-\epsilon z},\qquad
\text{with }h_0(z)e^{\gamma z} \sim A z^\nu,
\end{equation}
and it is clear that $\Psi_1(\gamma-\epsilon)$ is singular at $\epsilon=0$:
for $\nu\ge-1$, one has $\Psi_1(\gamma)=\infty$; for $\nu\in[-2,-1)$, then
$\Psi_1(\gamma)$ is finite but $\Psi_1'(\gamma)$ is infinite, etc.
In fact, still assuming~\eqref{againh0}, one can show that
\begin{equation}
\FST_\epsilon\big[\Psi_1(\gamma-\epsilon)\big]=
\left[\setlength{\arraycolsep}{-1pt}\begin{array}{l}\text{\small First singular term}\\
\text{\small in an
$\epsilon$ expansion}
\\\text{\small of $\Psi_1(\gamma-\epsilon)$}\end{array}\right]=
A
\begin{cases}
\Gamma(\nu+1)\epsilon^{-\nu-1}& \text{if $\nu\not\in\{-1,-2,-3,\ldots\}$},\\[1ex]
\dfrac{(-)^\nu\epsilon^{-\nu-1}\log\epsilon}{(-\nu-1)!}& \text{if $\nu\in\{-1,-2,-3,\ldots\}$}.
\end{cases}
\label{FSTPsi1}
\end{equation}
For example, when $\nu=-2.9$, we would write
$\FST_\epsilon[\Psi_1(\gamma-\epsilon)]=A\Gamma(-1.9)\epsilon^{1.9}$,
meaning that
$\Psi_1(\gamma-\epsilon)=\Psi_1(\gamma)-\Psi_1'(\gamma)\epsilon +
A\Gamma(-1.9)\epsilon^{1.9}+o(\epsilon^{1.9})$.
In general, the small $\epsilon$ expansion of
$\Psi_1(\gamma-\epsilon)$ starts like some polynomial in $\epsilon$ and,
then, the
first singular term is given by \eqref{FSTPsi1}.

One can understand \eqref{FSTPsi1}  by comparing $\Psi_1$ with 
the following functions of $\epsilon>0$, which
 can be written as an analytic function plus
one singular term:
\begin{equation}
\int_1^\infty\diffd z\, z^\nu e^{-\epsilon z}
=\big[\text{analytic function of $\epsilon$}\big]+
\begin{cases}
\Gamma(\nu+1)\epsilon^{-\nu-1}& \text{if $\nu\not\in\{-1,-2,-3,\ldots\}$}
,\\[1ex]
\dfrac{(-)^\nu\epsilon^{-\nu-1}\log\epsilon}{(-\nu-1)!}& \text{if
$\nu\in\{-1,-2,-3,\ldots\}$}
.
\end{cases}
\label{FST0}
\end{equation}
For $\nu>-1$, obtaining~\eqref{FST0} is easy: the analytic function is
simply $-\int_0^1\diffd z\,z^\nu e^{-\epsilon z}$. For $\nu=-1$, a similar argument
holds
after an integration by parts of $1/z$. For $\nu<-1$, one simply needs to 
integrate \eqref{FST0} with $\nu\ge-1$ over $\epsilon$ as many times as
needed. Note that one could change the lower bound of the integral in the
left hand side of \eqref{FST0} to any positive value without changing the
right hand side, as the nature of the singularity is governed by the
large $z$ regime.

\subsubsection*{The singularity in $\Psi_2$}

We now turn to writing the singularity in $\Psi_2$. We define
$\delta_t=o(t)$ as the sublinear correction in the position:
\begin{equation}
\mu_t = v t +\delta_t.
\label{mudelta}
\end{equation}
We need to consider two cases.
\begin{itemize}
\item If $\gamma<1$, then $\lambda=\gamma$, $v=\gamma+\gamma^{-1}$, and
$\Psi_2(r)$ is singular at $r=\gamma$. Using \eqref{mudelta} in
\eqref{defPsi} we obtain
\begin{equation}
\Psi_2(\gamma-\epsilon)=\int_0^\infty\diffd t\, e^{-\epsilon
(\gamma^{-1}-\gamma+\epsilon) t+(\gamma -\epsilon)\delta_t }
\approx
\int_0^\infty\diffd t\, e^{-\epsilon
(\gamma^{-1}-\gamma) t+\gamma \delta_t }
\quad\text{for $\epsilon>0$}.
\label{Psi2eps}
\end{equation}
It is then clear that, by taking $\delta_t$ logarithmic in $t$ for large
$t$, one recovers the same kind of integrals as in \eqref{FST0} with
$\epsilon$ replaced by $(\gamma^{-1}-\gamma)\epsilon$, and one
will be able to easily match the singularities with \eqref{FSTPsi1}. This
is done in detail in Section~\ref{sec>2}.
\item If $\gamma\ge1$, then $\lambda=1$ and $v=2$. The function $\Psi_2(r)$
is singular at $r=1$ if $\gamma=1$ and unexpectedly analytic at $r=1$ if
$\gamma>1$. With the form \eqref{mudelta}, one gets
\begin{equation}
\Psi_2(1-\epsilon)=\int_0^\infty\diffd t\,e^{-\epsilon^2
t+(1-\epsilon)\delta_t}
\quad\text{for $\epsilon>0$}.
\label{Psi2eps2}
\end{equation}
Again, by choosing $\delta_t$ logarithmic in $t$, one recovers the same
kind of integrals as in \eqref{FST0} but with $\epsilon$ replaced by
$\epsilon^2$. As shown below, this difference is important. The matching
of singularities with $\Psi_1$ is explained in Section~\ref{sec=2}.
\end{itemize}

\subsubsection*{Other useful identities}

We need in Sections~\ref{sec>2} and \ref{sec=2} some generalisations of \eqref{FST0} which we now
enumerate.
By taking the derivative of \eqref{FST0} with respect to
$\nu$ when $\nu\not\in\{-1,-2,-3,\ldots\}$, one gets
\begin{equation}
\int_1^\infty\diffd z\, (\log z) z^\nu e^{-\epsilon z}
=\big[\text{analytic function of $\epsilon$}\big]+
\Big[\Gamma(\nu+1)(-\log\epsilon)+\Gamma'(\nu+1)\Big]\epsilon^{-\nu-1}.
\label{derivlog}
\end{equation}
Let us write more explicitly the ones we actually use:
\begin{align}
\int_1^\infty\diffd z\, (\log z) z^{-\frac32} e^{-\epsilon z}
&=\big[\text{analytic function of $\epsilon$}\big]
+2\sqrt{\pi\epsilon}\big[\log\epsilon+\text{Cst}\big],
\label{FST1-1.5}
\\
\int_1^\infty\diffd z\, (\log z) z^{-\frac52} e^{-\epsilon z}
&=\big[\text{analytic function of $\epsilon$}\big]
-\frac43\sqrt{\pi}\epsilon^{\frac32}\big[\log\epsilon+\text{Cst}\big].
\label{FST1-2.5}
\end{align}
(The constants in the two lines are different.)
By taking the derivative of \eqref{derivlog} with respect to $\nu$, an
extra $\log z$ term appears in the integral and one obtains an expression
for $\int_1^\infty\diffd
z (\log z)^2z^\nu e^{-\epsilon z}$. We only need the case $\nu=-\frac32$,
which is given by
\begin{equation}
\int_1^\infty\diffd z\, (\log z)^2 z^{-\frac32} e^{-\epsilon z}
=\big[\text{analytic function of $\epsilon$}\big]
-2\sqrt{\pi\epsilon}\big[\log^2\epsilon-(4-4\log2-2\gamma_E)\log\epsilon+\text{Cst}\big]
\label{FST2-1.5}
\end{equation}
Taking $\nu=-2-u$ with $u>0$ in \eqref{FST0} leads, after a small $u$
expansion, to
\begin{equation}
\int_1^\infty\diffd z\, \frac{\log z}{z^2}e^{-\epsilon z}
=1-\epsilon\bigg(\frac{\log^2\epsilon}2-(1-\gamma_E)\log\epsilon+\text{Cst}\bigg)+\mathcal
O(\epsilon^2).
\label{FST1-2}
\end{equation}

\subsection{Sublinear terms in the position when
\texorpdfstring{$v>2$}{v>2}}
\label{sec>2}

We assume that
$h_0$ is of the form \eqref{againh0}: 
$h_0(x)\sim A x^\nu e^{-\gamma x}$. The first
singular term of $\Psi_1$ is given by \eqref{FSTPsi1} and
$\Psi_2(\gamma-\epsilon)$ is given by \eqref{Psi2eps}, where
$\delta_t=\mu_t -v t$.
It is easy to see that the singularities in $\Psi_2$ and $\Psi_1$ match
if
\begin{equation}
\delta_t=\frac\nu\gamma\log t +a+o(1)
\qquad\text{as $t\to\infty$},
\label{delta2}
\end{equation}
Indeed, comparing \eqref{Psi2eps} to \eqref{FST0} leads to
\begin{equation}
\FST_\epsilon\big[\Psi_2(\gamma-\epsilon)\big]=
\begin{cases}
e^{\gamma a}\Gamma(\nu+1)\big[(\gamma^{-1}-\gamma)\epsilon\big]^{-\nu-1}&
\text{if $\nu\not\in\{-1,-2,-3,\ldots\}$ ,}\\[1ex]
e^{\gamma
a}\dfrac{(-)^\nu\big[(\gamma^{-1}-\gamma)\epsilon\big]^{-\nu-1}\log\epsilon}{(-\nu-1)!}&
\text{if $\nu\in\{-1,-2,-3,\ldots\}$.}
\end{cases}
\label{FSTPsi2}
\end{equation}
The relation \eqref{magic2} yields that
\begin{equation}\label{sing match}
\FST_\epsilon\big[\Psi_1(\gamma-\epsilon)\big]=(\beta+\frac \alpha \gamma) \FST_\epsilon\big[\Psi_2(\gamma-\epsilon)\big].
\end{equation}
Thus,  one must choose $a$ such that
\begin{equation}
A=\left(\beta+\frac\alpha\gamma\right)e^{\gamma
a}(\gamma^{-1}-\gamma)^{-\nu-1}.
\end{equation}
Finally,
\begin{equation}
\mu_t=v t +\frac\nu\gamma\log t +\underbrace{\frac1\gamma\log \frac
{A\gamma(\gamma^{-1}-\gamma)^{\nu+1}}{\alpha+\beta\gamma}}_a+o(1).
\end{equation}

\subsection{Sublinear terms in the position when
\texorpdfstring{$v=2$}{v=2}}
\label{sec=2}

The case $v=2$ corresponds to $\lambda=1$. Either $\gamma>1$, and both
$\Psi_1$ and $\Psi_2$ are analytic around $r=1$, or $\gamma=1$ and
they have matching singularities.
$\Psi_2(1-\epsilon)$ is given by \eqref{Psi2eps2} where $\delta_t=\mu_t-2t$.
If one chooses $\delta_t$ logarithmic in $t$
\begin{equation}
\delta_t=\xi\log t+a+o(1),
\label{deltat2}
\end{equation}
then, using \eqref{FST0} with $\epsilon$ replaced by $\epsilon^2$, one gets
from \eqref{Psi2eps2}
\begin{equation}
\FST_{\epsilon^2}\big[\Psi_2(1-\epsilon)\big]=
\begin{cases}
e^a\Gamma(\xi+1)(\epsilon^2)^{-\xi-1}&
\text{if $\xi\not\in\{-1,-2,-3,\ldots\}$}
,\\[1.5ex]
e^a\dfrac{(-)^\xi(\epsilon^2)^{-\xi-1}\log(\epsilon^2)}{(-\xi-1)!}
&\text{if $\xi\in\{-1,-2,-3,\ldots\}$.}
\end{cases}
\label{FSTcrita}
\end{equation}
Observe that this is the first singular term for an expansion in powers of
$\epsilon^2$. In an expansion in powers of $\epsilon$, the term above is
not singular for $\xi\in\{-\frac32,-\frac52,-\frac72,\ldots\}$, and one 
concludes that, in an expansion in powers of $\epsilon$,
\begin{equation}
\FST_{\epsilon}\big[\Psi_2(1-\epsilon)\big]=
\begin{cases}
e^a\Gamma(\xi+1)\epsilon^{-2\xi-2}&\text{if 
$\xi\not\in\{-1,-\frac32,-2,-\frac52,-3,-\frac72,\ldots\}$},\\[1.5ex]
(-)^\xi 2 e^a\dfrac{\epsilon^{-2\xi-2}\log\epsilon}{(-\xi-1)!}
&\text{if $\xi\in\{-1,-2,-3,\ldots\}$},\\[1.5ex]
o\big(\epsilon^{-2\xi-2}\big)&\text{if
$\xi\in\{-\frac32,-\frac52,-\frac72,\ldots\}$.}
\end{cases}
\label{FSTcrit}
\end{equation}

Let us assume that $h_0$ is of the form \eqref{againh0} with $\gamma=1$:
$h_0(x)\sim A x^\nu e^{-x}$. The first singular term in $\epsilon$
must be, from \eqref{FSTPsi1}, either in $\epsilon^{-\nu -1}$
or in $\epsilon^{-\nu-1}\log \epsilon$.
For example, if $\nu=-1.8$, then the singularity is $\epsilon^{0.8}$ and
\eqref{FSTcrit} leads to $\xi=-1.4$. On the other hand, if $\nu=-2.2$, the
singularity is $\epsilon^{1.2}$ and \eqref{FSTcrit} gives two possible
solutions: either $\xi=-1.6$ or $\xi=-\frac32$. Thus, we see that the case
$\nu\ge-2$ (where there is no ambiguity) is simpler than the case
$\nu<-2$ (where one needs to determine the correct solution amongst several
possibilities).

\subsubsection*{The case $\gamma=1$ and $\nu\ge-2$}

When $\nu>-2$, one can
match unambiguously \eqref{FSTPsi1} and \eqref{FSTcrit} and one finds that
$\xi=(\nu-1)/2$. Putting aside the case $\nu=-1$ for now, one finds 
\begin{equation}
A\Gamma(\nu+1)=(\alpha+\beta)e^a\Gamma\left(\dfrac{\nu+1}2\right)\quad\text{for
$\nu>-2$ and $\nu\ne-1$,}
\end{equation}
and finally
\begin{equation}
\mu_t=2t+\frac{\nu-1}{2}\log
t+\underbrace{\log\frac{A\Gamma(\nu+1)}{(\alpha+\beta)\Gamma[(\nu+1)/2]}}_a
+o(1)\qquad\text{for $\nu>-2$ and $\nu\ne-1$.}
\label{nu>-2}
\end{equation}
For $\nu=-1$, the singularity is of order $\log\epsilon$ and a simple
matching gives
\begin{equation}
\mu_t=2t-\log
t+\underbrace{\log\frac{2A}{\alpha+\beta}}_a
+o(1)\qquad\text{for $\nu=-1$.}
\end{equation}
The case $\nu=-2$ is slightly more problematic because the first singular
term in $\Psi_1$ is $A\epsilon\log\epsilon$ which cannot be obtained from
\eqref{FSTcrit}. The correction to the position is not of the
form~\eqref{deltat2}, but one can check that in fact $\delta_t=-\frac32\log
t+\log\log t +a +o(1)$ matches the singularity and, finally, using
\eqref{FST1-1.5},
\begin{equation}
\mu_t=2t-\frac32\log t+\log\log t + \underbrace{\log\frac
A{(\alpha+\beta)4\sqrt\pi}}_a+o(1)
\qquad\text{for $\nu=-2$.}
\end{equation}

\subsubsection*{The leading term when $\gamma=1$ and $\nu<-2$}

When $\nu<-2$, the first singular term in $\Psi_1$ is small compared to
$\epsilon$, see \eqref{FSTPsi1}, and there are several ways of matching
such a singular term from \eqref{FSTcrit}. For instance, if $\nu=-2.2$, 
the singular term is $\epsilon^{1.2}$ and, from \eqref{FSTcrit}, one has either
$\xi=(\nu-1)/2=-1.6$  or $\xi=-3/2$. If $\nu=-4.2$, one could either have 
$\xi=-2.6$ or $\xi=-3/2$ or $\xi=-5/2$, etc.

To resolve this difficulty, we now argue
that there exists a constant $C$ such that, for any initial
condition $h_0$, one has 
\begin{equation}
 \delta_t+(3/2)\log t \ge C\qquad\text{for $t$ large
enough}.
\label{bd}
\end{equation}
This will imply that $\xi \ge -3/2$, always, and thus that we have
$\xi=-3/2$ when $\nu<-2$. In fact, more
generally, $\xi=-3/2$
for any initial condition 
such that $\FST_\epsilon[\Psi_1(1-\epsilon)]=o(\epsilon)$ or, equivalently,
such that
$\int\diffd x\,h_0(x)x e^x<\infty$:
\begin{equation}
\mu_t=2t-\frac32\log t + a +o(1)\qquad\text{if and only if }\int\diffd x\,h_0(x)x
e^x<\infty,
\end{equation}
for some constant $a$.

We now turn to showing \eqref{bd}. We only need to consider the case
($\alpha=1$, $\beta=0$) because of the mapping explained in
Section~\ref{equiv}. With the construction of a solution
to the ($\alpha=1$, $\beta=0$) explained in Section~\ref{const}, it is
clear that there is a comparison principle:
for two ordered initial conditions $h_0(x)\le \tilde h_0(x)$, one
must have $\mu_t\le\tilde\mu_t$ at all times where ($\mu_t$ resp.\@
$\tilde\mu_t$) is the position of the solution with initial condition $h_0$
(resp.\@ $\tilde h_0$). This implies that it is sufficient to show
\eqref{bd} for the initial condition $h_0=0$.

When $h_0=0$, the main relation~\eqref{magic2} reduces to
$\Psi_2(1-\epsilon)=1$ for $\epsilon>0$. Writing as usual
$\mu_t=2t+\delta_t$, we expand \eqref{Psi2eps2} up to
order $o(\epsilon)$. 
\begin{equation}\label{eric's expansion}
 \Psi_2(1-\epsilon) =\int_0^\infty \diffd t \, e^{-\epsilon^2t +(1-\epsilon)\delta_t}
= \int_0^\infty \diffd t \, e^{\delta_t}e^{-\epsilon^2 t} -\epsilon\int_0^\infty \diffd t\, \delta_t e^{\delta_t} +o(\epsilon).
\end{equation}

The behaviour of the first term in the right-hand-side depends on $\exp (\delta_t)$. From \eqref{FST0}:
\begin{itemize}
\item If $\exp(\delta_t) \sim e^a t^{-3/2}$ then $\int_0^\infty \diffd t \, e^{\delta_t}e^{-\epsilon^2 t}  = \int_0^\infty \diffd t \, e^{\delta_t} - \epsilon 2e^a\sqrt \pi +o(\epsilon)$.

\item If $\exp(\delta_t) =o(t^{-3/2})$, for instance if $\delta_t =\xi \log
t +a+o(1)$ with $\xi<-3/2$, then one finds $\int_0^\infty \diffd t \, e^{\delta_t}e^{-\epsilon^2 t}  = \int_0^\infty \diffd t \, e^{\delta_t}+o(\epsilon)$.

\item Finally, if $\exp(\delta_t)\gg t^{-3/2}$, for instance if $\delta_t =\xi \log t +a+o(1)$ with $\xi>-3/2$, then $\int_0^\infty \diffd t \, e^{\delta_t}e^{-\epsilon^2 t}  = \int_0^\infty \diffd t \, e^{\delta_t}+R(\epsilon)$ with $R(\epsilon)\gg \epsilon $.
\end{itemize}

The initial condition
$h_0=0$ is of course below the travelling wave, and the position of the
front started from the travelling wave is exactly $2t$. By using again
the comparison principle, we conclude that the position of the front with
$h_0=0$ is such that $\delta_t\le0$ at all times. 
Thus, $\int_0^\infty \diffd t\, \delta_t e^{\delta_t} <0$ and the only way that 
$\Psi_2(1-\epsilon)=1$ is that 
\begin{equation}
\exp(\delta_t) \sim e^a t^{-3/2}\quad\text{with}\quad 2e^a \sqrt\pi=-\int_0^\infty \diffd t\, \delta_t e^{\delta_t}
\end{equation}
This concludes the argument.
We were not able to find a simpler expression for $a$.

\subsubsection*{Vanishing corrections when $\gamma=1$ and $\nu<-2$}

We still consider the case where $\int\diffd x \, h_0(x) x e^x<\infty$ or,
when considering only initial conditions of the form $h_0(x)\sim A x^\nu
e^{-x}$, the case $\nu<-2$. From the previous argument, we must have
\begin{equation}
\delta_t = -\frac32\log t + a + q_t\qquad\text{with }q_t=o(1).
\label{rt}
\end{equation}
In \eqref{Psi2eps2},
\begin{equation}
\Psi_2(1-\epsilon)=e^{a(1-\epsilon)}
\int_1^\infty\diffd t\, \frac{e^{-\epsilon^2
t}}{t^{3/2}} e^{\epsilon\frac32\log t +q_t -\epsilon q_t}+f(\epsilon),
\label{psir}
\end{equation}
where $f(\epsilon)$, which represents the integral from 0 to 1, has
no singularity in a small $\epsilon$ expansion.

The singularity of $\Psi_2(1-\epsilon)$ at $\epsilon=0$
is dominated by the large $t$ decay.
The second exponential in \eqref{psir} can be
expanded; the leading term in this expansion gives
$\int_1^\infty\diffd t\, e^{-\epsilon^2t}/t^{3/2}$,
which is not singular in $\epsilon$. 
The next term is either
\begin{equation}
\FST_\epsilon\left[\int_1^\infty\diffd t\,
\frac{e^{-\epsilon^2
t}}{t^{3/2}}\epsilon\frac32\log t\right]
= 6\sqrt\pi\epsilon^2\log\epsilon,
\label{82}
\end{equation}
see
\eqref{FST1-1.5}, 
or
\begin{equation}
\FST_\epsilon\left[\int_1^\infty\diffd t\,
\frac{e^{-\epsilon^2
t}}{t^{3/2}}q_t\right],
\label{83}
\end{equation}
or the sum of the two if \eqref{83} is also of order
$\epsilon^2\log\epsilon$, or a higher order term is the sum cancels.

From \eqref{FST0}, the quantity \eqref{83} is of order
$\epsilon^2\log\epsilon$ when $q_t$ decays as $1/\sqrt t$:
\begin{equation}
\FST_\epsilon\left[\int_1^\infty\diffd t\,
\frac{e^{-\epsilon^2
t}}{t^{3/2}}\frac1{\sqrt t}\right]= 2\epsilon^2\log\epsilon.
\label{84}
\end{equation}
We are now ready to match the singularities of $\Psi_1$ and $\Psi_2$. Recall that for
 an initial
condition $h_0(x)\sim A x^\nu e^{-x}$, the first singular term in $\Psi_1$
is given by \eqref{FSTPsi1} and is of order $\epsilon^{-\nu-1}$.
We consider three cases.
\begin{itemize}
\item{When $\nu\in(-3,-2)$,} the singularity in $\Psi_1$ is between $\epsilon$ and
$\epsilon^2$ and can only be matched by the $q_t$ term \eqref{83}. One must
choose $q_t\sim -b t^{1+\nu/2}$ which leads to
\begin{equation}
A\Gamma(\nu+1) =-(\alpha+\beta)e^a b \Gamma\Big(\frac{\nu+1}2\Big),
\end{equation}
and finally
\begin{equation}
\mu_t=2t-\frac32\log
t + a -\underbrace{\frac{-Ae^{-a}\Gamma(\nu+1)}{\Gamma\big[(\nu+1)/2\big](\alpha+\beta)}}_{b}
 t^{1+\frac\nu2}+o(t^{1+\frac\nu2})\qquad\text{for $\nu\in(-3,-2)$}.
\end{equation}
\item{When $\nu=-3$,} the singular term in $\Psi_1$ is
$-(A/2)\epsilon^2\log\epsilon$. One needs to take $q_t$ of the form
$q_t\sim -b t^{-1/2}$ and both \eqref{82} and \eqref{84} contribute.
Matching singularities gives
\begin{equation}
-\frac A2=(\alpha+\beta)e^a\Big[6\sqrt\pi-2b\Big],
\end{equation}
and finally
\begin{equation}
\mu_t=2t-\frac32\log
t + a -\underbrace{\Big[\frac{Ae^{-a}}{4(\alpha+\beta)}+3\sqrt\pi\Big]}_{b}
 t^{-1/2}+o(t^{-1/2})\qquad\text{for $\nu=-3$}.
\end{equation}

\item{When $\nu<-3$,}
there cannot exist
a $\epsilon^2\log\epsilon$ singularity in $\Psi_2$ and the terms \eqref{82}
and \eqref{83} must cancel. This leads to
\begin{equation}
\mu_t=2t-\frac32\log
t + a -\frac{3\sqrt\pi}{\sqrt t}
+o(t^{-1/2})
\qquad\text{for $\nu<-3$}.
\label{87}
\end{equation}
\end{itemize}

For general initial condition (not necessarily such that $h_0\sim A x^\nu
e^{-x}$), a necessary and sufficient condition to have the expansion \eqref{87} is simply that
the first singular term in $\Psi_1(1-\epsilon)$ is smaller than
$\epsilon^2\log\epsilon$:
\begin{equation}
\mu_t=2t-\frac32\log
t + a -\frac{3\sqrt\pi}{\sqrt t}
+o(t^{-1/2})\qquad\text{if and only if
}\FST_\epsilon[\Psi_1(1-\epsilon)]=o(\epsilon^2\log\epsilon).
\label{truecond1}
\end{equation}
In particular, the condition in \eqref{pull1}, which is equivalent to
$\FST_\epsilon[\Psi_1(1-\epsilon)]=o(\epsilon^2)$, is sufficient.

\subsubsection*{Second order vanishing corrections when $\gamma=1$ and $\nu<-3$}

We only consider the case where
\eqref{truecond1} holds; 
when considering initial conditions of the form $h_0(x)\sim A x^\nu
e^{-x}$, this corresponds to $\nu<-3$. We write
\begin{equation}
\delta_t=-\frac32\log t+a-3\frac{\sqrt\pi}{\sqrt t} + s_t\qquad\text{with
}s_t=o(t^{-1/2}),
\end{equation}
and, as in \eqref{psir},
\begin{equation}
\Psi_2(1-\epsilon)=e^{a(1-\epsilon)}
\int_1^\infty\diffd t\, \frac{e^{-\epsilon^2
t}}{t^{3/2}} e^{\epsilon\frac32\log t - 3\sqrt{\pi/t} +s_t + \epsilon
\big(3\sqrt{\pi/t} -s_t \big)}+f(\epsilon).
\label{psir2}
\end{equation}
We expand the last exponential in \eqref{psir2}, keeping all the terms that
lead to singularities larger than $\epsilon^3$. The terms with
$\epsilon\frac32\log t$ and
$-3\sqrt{\pi/t}$ have cancelling singularities, so there  remains
\begin{equation}
\FST_\epsilon\left[\int_1^\infty\diffd t\,
\frac{e^{-\epsilon^2
t}}{t^{3/2}}s_t\right],
\label{98}
\end{equation}
which is not yet known, and
\begin{equation}
\FST_\epsilon\left[\int_1^\infty\diffd t\,
\frac{e^{-\epsilon^2
t}}{t^{3/2}}\bigg[\frac12 \bigg(\epsilon\frac32\log
t-\frac{3\sqrt\pi}{\sqrt
t}\bigg)^2+\epsilon\frac{3\sqrt\pi}{\sqrt t}\bigg]\right]=(15-18\log2)\sqrt\pi \epsilon^3\log\epsilon,
\label{99}
\end{equation}
which we computed using \eqref{FST0}, \eqref{FST2-1.5} and \eqref{FST1-2}.

The argument is then the same as before and, for an  initial condition
$h_0(x)\sim A x^\nu e^{-x}$,  one needs to consider three
cases.
\begin{itemize}
\item When $\nu\in(-4,-3)$, the singularity in
$\Psi_1$ is between $\epsilon^2$ and $\epsilon^3$  and must be matched by
\eqref{98} because \eqref{99} is too small. This leads to taking $s_t\sim
-b t^{1+\nu/2}$ and finally
\begin{equation}
\mu_t=2t-\frac32\log t + a -\frac{3\sqrt\pi}{\sqrt
t}-\underbrace{\frac{-Ae^{-a}\Gamma(\nu+1)}{\Gamma\big[(\nu+1)/2\big](\alpha+\beta)}}_b t^{1+\frac\nu2}+o(t^{1+\frac\nu2})\qquad\text{for
$\nu\in(-4,-3)$}.
\end{equation}
\item When $\nu=-4$, the singularity in $\Psi_1$ is
$A\epsilon^3(\log\epsilon)/6$. The term \eqref{99} contributes, as well as
\eqref{98} with $s_t\sim b(\log t)/t$ using \eqref{FST1-2.5}:
\begin{equation}
\mu_t=2t-\frac32\log t + a -\frac{3\sqrt\pi}{\sqrt
t}+\underbrace{\left(\frac98\big(5-6\log
2\big)-\frac{Ae^{-a}}{16\sqrt\pi(\alpha+\beta)}\right)}_b
\frac{\log t}t+o\Big(\frac{\log t}t\Big)\quad\text{for
$\nu=-4$}.
\end{equation}
\item When $\nu<-4$,
there cannot exist
a $\epsilon^3\log\epsilon$ singularity in $\Psi_2$ and the terms \eqref{98}
and \eqref{99} must cancel. This leads to $s_t$ of order $(\log t)/t$ and
\begin{equation}
\mu_t=2t-\frac32\log t + a -\frac{3\sqrt\pi}{\sqrt
t}
+\frac98\big(5-6\log
2\big)
\frac{\log t}t+o\Big(\frac{\log t}t\Big)\quad\text{for $\nu<-4$}.
\label{95}
\end{equation}
\end{itemize}
For general initial condition (not necessarily such that $h_0\sim A x^\nu
,e^{-x}$), a necessary and sufficient condition to have the expansion
\eqref{95} is simply that
the first singular term in $\Psi_1(1-\epsilon)$ is smaller than
$\epsilon^3\log\epsilon$:
\begin{equation}
\mu_t=2t-\frac32\log
t + a -\frac{3\sqrt\pi}{\sqrt t}
+\frac98\big(5-6\log
2\big)
\frac{\log t}t+o\Big(\frac{\log t}t\Big)
\quad\text{iff
}\FST_\epsilon[\Psi_1(1-\epsilon)]=o(\epsilon^3\log\epsilon).
\label{truecond2}
\end{equation}
In particular, the condition in \eqref{pull1}, which is equivalent to
$\FST_\epsilon[\Psi_1(1-\epsilon)]=o(\epsilon^3)$, is sufficient.

\section{What happens when the front does not go to a travelling wave}\label{sec:Munier}

In Section~\ref{sec:rem}, we established that the solution to \eqref{ours}
cannot converge to a travelling wave unless condition \eqref{condM0} is
satisfied. In this section, we investigate briefly what happens when
\eqref{condM0} is not met.

Let $h(x,t)$ be the solution to \eqref{ours} with initial condition
$h_0$. (As usual, we assume that $h_0\ge0$ decays exponentially fast at
infinity.) To simplify the discussion, we only consider the case
($\alpha=0$, $\beta=1$).
As in \eqref{eta0h0}, let us introduce $\eta_0(x)$ by
\begin{equation}
\eta_0(x)=1-\int_0^x\diffd z\, h_0(z),
\end{equation}
and recall from \eqref{etatoh} that $h(x,t)$ can be written as
\begin{equation}
h(x,t)=-\partial_x \eta(x,t),
\end{equation}
where $\eta(x,t)$ is the solution to \eqref{ours} with parameters
($\alpha=1$, $\beta=0$) and initial condition $\eta_0$.

The total mass  in the system at time~$t$ is given by $g(0,t)$
and its evolution is simply given by \eqref{gsp}:
\begin{equation}
g(0,t)=\int_0^\infty\diffd z\, h(\mu_t+z,t)
=
1 + \Big(\int_0^\infty \diffd z\,h_0(z)-1\Big)e^t.
\end{equation}

There are three cases to consider
in order to understand the evolution of the front $h$:
\begin{description}
\item[If $\displaystyle\int_0^\infty \diffd z\,h_0(z)=1$.]\ \\[1ex]
Then \eqref{condM0} holds. The 
function $\eta_0(x)=\int_x^\infty \diffd z\,h_0(z)$ is a decreasing
function decaying exponentially fast at infinity. The front $\eta(x,t)$
converges to a travelling wave, and so does $h(x,t)$. Notice that the mass
$g(0,t)$ is equal to 1 at all time $t$.

\item[If  $\displaystyle\int_0^\infty \diffd z\,h_0(z)<1$.]\ \\[1ex]
Then \eqref{condM0} does not hold. The mass $g(0,t)$ reaches zero at some finite
time $t_c$. As $h(x,t)$ is non-negative, this implies that
$h(\mu_t+z,t)\to0$ for all $z>0$ as $t\to t_c$; the front disappears in
finite time, and the solution to \eqref{ours} does not exist for $t>t_c$.
This is easy to understand by considering $\eta(x,t)$; the initial
condition $\eta_0(x)$ is a decreasing function bounded away from zero:
$\eta_0(\infty)=1-\int\diffd z \,h_0(z)>0$. At all times, $\eta(x,t)$ is
a decreasing function of $x$ with $\eta(\infty,t)=\big(1-\int\diffd
z \,h_0(z)\big)e^t=1-g(0,t)\to 1$  as $t\to t_c$. Thus, $\eta(x,t)\to 1$
uniformly while $\mu_t$  diverges.

\item[If  $\displaystyle\int_0^\infty \diffd z\,h_0(z)>1$.]\ \\[1ex]
Then \eqref{condM0} does not hold. The mass $g(0,t)$ diverges exponentially.
We argue below that the position $\mu_t$ of the boundary runs to the left
with velocity 2 and that it converges to the pseudo-travelling wave
$z e^z$:
\begin{equation}
h(\mu_t+z,t)\to z e^z\qquad\text{with}\qquad-\mu_t=2t-\frac32\log
t+a+3\frac{\sqrt\pi}{\sqrt t}+ o(t^{-\frac12}).
\label{reversed}
\end{equation} 
(In Section~\ref{sec:TW}, we insisted that a proper travelling wave
$\omega_v$ must be positive and
satisfy $\omega_v(\infty)=0$, so in that sense $ze^z$ is not really a
travelling wave.)
We wrote $-\mu_t$ rather than $\mu_t$
in \eqref{reversed} to have the usual signs for the
velocity and logarithmic correction. Notice that the $1/\sqrt t$ correction
has the same coefficient with an opposite sign as the Ebert and van
Saarloos correction for the position of the Fisher-KPP front.
\end{description}
An intuitive way to visualize these three cases is the following:
as $\alpha=0$, the boundary at $\mu_t$ is an absorbing boundary.
If it were not moving ($\mu_t=0$ $\forall t$), the front $h$ would grow
and so would the slope at $\mu_t$.
In order to fix $\beta=\partial_xh(\mu_t,t)=1$, we must prevent this growth.  There are two
strategies: either $\mu_t$ moves to the right to prevent the front from
growing, or $\mu_t$ moves to the left to ``escape'' the ever-growing
front and to find a region where the slope is not yet large. The first two cases
above correspond to the first strategy, while the third case corresponds to
the second strategy.

It would be sufficient to assume that $h(\mu_t+z,t)$ has a large time limit
to derive
\eqref{reversed} using the techniques developed in the present paper. To
make this section short and simple, we now make the stronger assumption that
$h(\mu_t+z,t)\to ze^z$ and $-\mu_t=2t+o(t)$, and we explain briefly
how the sub-linear terms in $\mu_t$ can be obtained.
The derivation of Section~\ref{sec:magic} still holds, except that
$g(r,t)$ has a long time limit only for $r<-1$. We conclude that
\begin{equation}
\Psi_1(r)=\Psi_2(r)\text{ for $r<-1$,\quad with }
\Psi_1(r)=\int_0^\infty \diffd z\, h_0(z)e^{rz},\
\Psi_2(r)= \int_0^\infty \diffd t\, e^{r\mu_t-(r^2+1)t}.
\end{equation}
(This is the same as~\eqref{defPsi} and~\eqref{magic2} 
with ($\alpha=0$, $\beta=1$), but with
the extra condition $r<-1$.)

Following Section~\ref{sec=2}, we write $\mu_t=-(2t+\delta_t)$ and
$r=-1-\epsilon$ and obtain
\begin{equation}
\Psi_2(-1-\epsilon)=\int_0^\infty\diffd t\, e^{-\epsilon^2
t + (1+\epsilon)\delta_t}.
\label{Psi2new}
\end{equation}
The right hand side is similar to \eqref{Psi2eps2}, but with the opposite
sign for the term $\epsilon\delta_t$. As in Section~\ref{sec=2}, the question
is how to choose $\delta_t$ in such a way that \eqref{Psi2new} has no
singularity as $\epsilon\to0^+$. By following the same line of argument as in
Section~\ref{sec=2}, one finds that 
$\delta_t=-\frac32\log t + a + 3\sqrt{\pi/t}+o(1/\sqrt t)$, which is the
same result as in Section~\ref{sec=2} except for the sign of the $1/\sqrt
t$ correction. The difference comes directly from the sign difference of the 
$\epsilon\delta_t$ term between
\eqref{Psi2eps2} and
\eqref{Psi2new}.

Remark that an expression similar to \eqref{reversed}, with its unexpected
sign in front of the $1/\sqrt t$ correction, has already been obtained in
\cite{MuellerMunier.2014}. In that paper, the authors study the typical
density of particles in a branching Brownian motion. The \emph{expected}
density of particles $\rho$ satisfies the linear equation
$\partial_t\rho=\partial_x^2\rho+\rho$, but $\rho$ does not represent well
the typical density of particles because of the effect of rare paths
leading to many particles. In \cite{MuellerMunier.2014} it was
proposed to rather consider  $\psi(x,t)$ defined as the expected
density of particles who never went further than some $\bar X_t$ from their
starting point. The equation followed by $\psi$ is then
$\partial_t\psi=\partial_x^2\psi+\psi$ for $|x|<\bar X_t$
and $\psi(\pm\bar X_t,t)=0$. In this view, $\bar X_t$ had to be determined in
a self consistent way by requiring that the density $\psi$ at a distance of
order 1 from $\pm \bar X_t$ is of order 1. This equation is very similar to
our problem \eqref{ours} with ($\alpha=0$, $\beta=1$) and, in fact, it
was found \cite{MuellerMunier.2014} that $\psi(\bar X_t-z,t)\propto
z e^z$ and
that $\bar X_t$ has the same asymptotics as $-\mu_t$ in \eqref{reversed}.

\section{Conclusion}

In this work we have studied the long time asymptotics of the solutions of
(\ref{ours}).  When the solution converges to a travelling wave solution,
we have obtained precise expressions \eqref{pull1}, \eqref{pull2},
\eqref{pushedresults}, \eqref{criticalresults} for the position of the
front. In the pulled case our linear problem \eqref{ours} reproduces the
known expected asymptotics for the Fisher-KPP like equations, including
Bramson's logarithmic shift~\cite{Bramson.1983} and the power law
correction predicted by Ebert and van Saarloos
\cite{EbertvanSaarloos.2000}, see \eqref{pull1}. Our analysis allowed us to
even predict a further logarithmic correction, see last line of
\eqref{pull1}, and this raises the question of the existence and of the
universality of a whole series of correction terms for travelling wave
equations in the Fisher-KPP class with fast enough decaying initial
conditions. For our linear problem (\ref{ours}), we could also analyse the
pushed case, see \eqref{pushedresults} and \eqref{criticalresults}.

Surprisingly all the cases could be analysed from a single compact
equation \eqref{main}. This equation relates the position $\mu_t$ at time
$t$ of the front to the initial condition $h_0(x)$, and the large time
asymptotics of $\mu_t$ can be obtained by matching the first singularity of
both sides of \eqref{main}.

As an illustration of our results, by choosing ($\alpha=1$, $\beta=0$)
and $h_0(x)=0$ we find \eqref{main} that the
position $\mu_t$ is implicitly given by the following integral equation:
\begin{equation}
\int_0^\infty \diffd t\, e^{r \mu_t - (1+r^2) t}  =1,\quad\text{for all $r<1$}. \label{main-bis}
\end{equation}
This  leads to the asymptotics predicted by
\eqref{pull1}:
\begin{equation}
\mu_t=2t-\frac32\log t+\cst-3\frac{\sqrt\pi}{\sqrt t}+
		\frac98(5-6\log2)\frac{\log t}t + \cdots
\label{lastasymp}
\end{equation}
It would certainly be interesting to develop a more direct approach to
equations of the type \eqref{main-bis} to extract 
asymptotics such as \eqref{lastasymp}.

A question which remains is to formulate the  precise conditions that the
initial condition $h_0$ should satisfy for the solution to converge to
a travelling wave and to analyse the general case when it does not.

\end{document}